\documentclass[journal,twoside,web]{ieeecolor}
\usepackage{generic}
\usepackage{cite}
\usepackage{amsmath,amssymb,amsfonts}
\usepackage{algorithm}
\usepackage{algorithmic}
\usepackage{graphicx}
\usepackage{textcomp}
\usepackage{multirow}

\begin{document}
\title{Omni 3D: BEOL-Compatible 3D Logic with \\Omnipresent Power, Signal, and Clock}

\author{Suhyeong Choi, \IEEEmembership{Graduate Student Member, IEEE}, Carlo Gilardi, \IEEEmembership{Member, IEEE}, \\Paul Gutwin, \IEEEmembership{Member, IEEE}, Robert M. Radway, \IEEEmembership{Graduate Student Member, IEEE}, \\Tathagata Srimani, \IEEEmembership{Member, IEEE}, and Subhasish Mitra, \IEEEmembership{Fellow, IEEE}
\thanks{This manuscript was drafted on September 9, 2024 as extension of~\cite{omni}.}
\thanks{Suhyeong Choi, Carlo Gilardi, and Robert M. Radway are with the Department of Electrical Engineering, Stanford University.}
\thanks{Paul Gutwin is with Logic Technology Development, Intel Corporation.}
\thanks{Tathagata Srimani is with the Department of Electrical and Computer Engineering, Carnegie Mellon University.}
\thanks{Subhasish Mitra is with the Department of Electrical Engineering and the Department of Computer Science, Stanford University.}
}

\maketitle

\begin{abstract}
This paper presents Omni 3D --- a 3D-stacked device architecture that is naturally enabled by back-end-of-line (BEOL)-compatible transistors. Omni 3D arbitrarily interleaves metal layers for both signal/power with FETs in 3D (i.e., nFETs and pFETs are stacked in 3D). Thus, signal/power routing layers have fine-grained, all-sided access to the FET active regions maximizing 3D standard cell design flexibility. This is in sharp contrast to approaches such as back-side power delivery networks (BSPDNs), complementary FETs (CFETs), and stacked FETs. Importantly, the routing flexibility of Omni 3D is enabled by double-side routing and an interleaved metal (IM) layer for inter- and intra-cell routing, respectively. In this work, we explore Omni 3D variants (e.g., both with and without the IM layer) and optimize these variants using a virtual-source BEOL-FET compact model. We establish a physical design flow that efficiently utilizes the double-side routing in Omni 3D and perform a thorough design-technology-co-optimization (DTCO) of Omni 3D device architecture on several design points. From our design flow, we project $2.0\times$ improvement in the energy-delay product and $1.5\times$ reduction in area compared to the state-of-the-art CFETs with BSPDNs.

\end{abstract}


\section{Introduction}
\IEEEPARstart{U}ltra-dense 3D integration of logic and memory (with 3D vertical connectivity  $\leq 100 nm$) promises significant improvements in computing energy efficiency and throughput, particularly for today’s challenging abundant-data applications~\cite{n3xt}. Today, we can realize such ultra-dense 3D systems using monolithic 3D integration of back-end-of-line (BEOL)-compatible logic and memory technologies (that are fabricated at temperatures $\leq 400\,^\circ \text{C}$ to prevent damage to upper BEOL metal interconnect layers). BEOL-compatible logic has been experimentally demonstrated using low-dimensional field-effect transistors (FETs), such as MoS$_2$ FETs and carbon nanotube FETs (CNFETs)~\cite{lowd, 2d, cnt}. Multiple gate geometries (e.g., back-gated, top-gated, and gate-all-around~\cite{back, top, gaa, barrier}) have been extensively studied for BEOL-compatible logic. However, device architectures that co-optimize 3D arrangement of BEOL-compatible FETs alongside metal connections have not been sufficiently explored.

\begin{figure}[!t]
\centering
\includegraphics[width=3.3in]{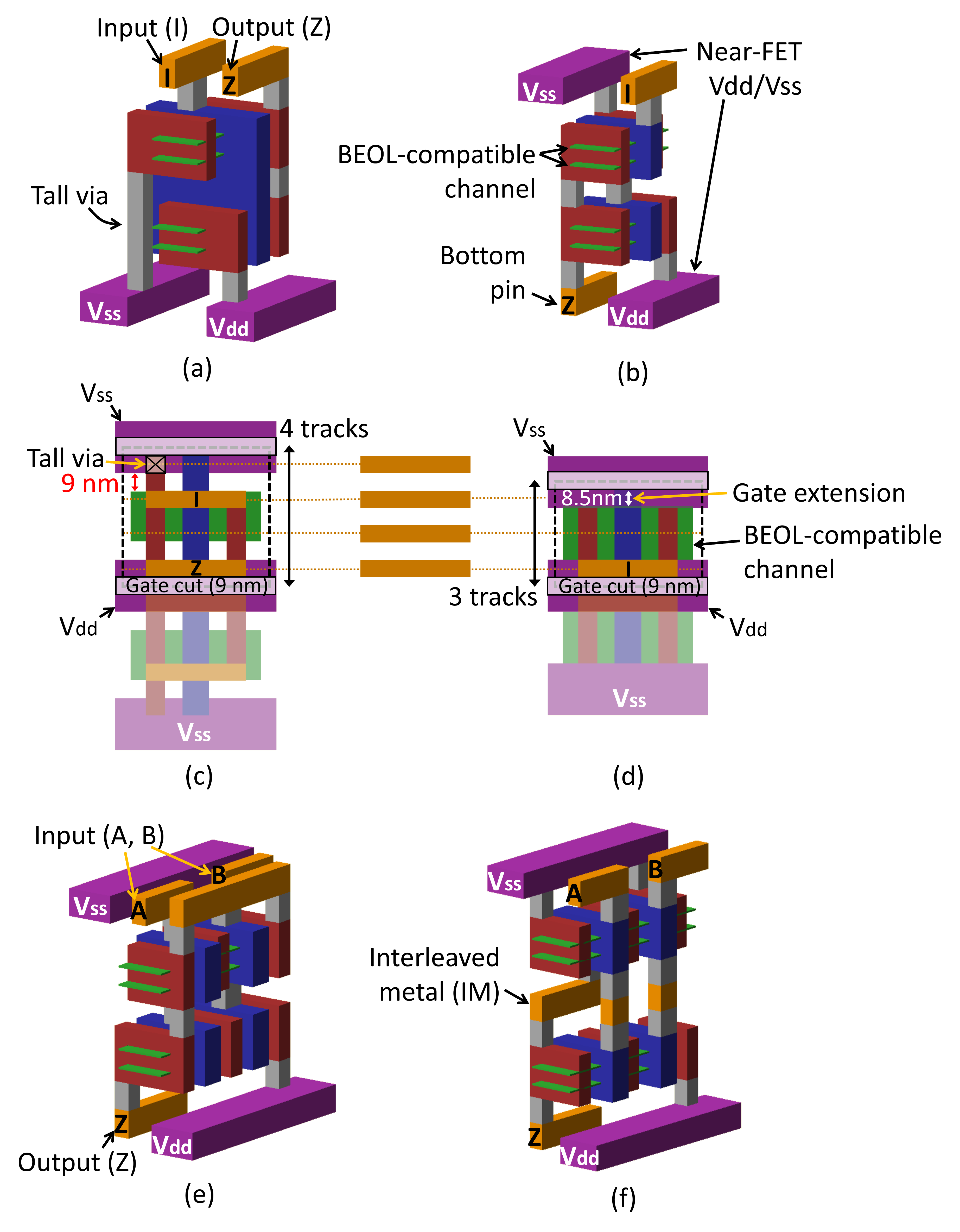}
\caption{Omni 3D features in contrast to CFET: 3D illustration of (a) a CFET inverter (INV) with BPR and (b) an Omni 3D INV with near-FET Vdd/Vss to avoid a tall via crossing the lower FET for upper FET power supply. A BEOL-compatible channel naturally enables the bottom pin in Omni 3D. Correspondingly, a top view of (c) 4-track CFET and (d) 3-track Omni 3D INVs with comparable channel width. CFET channel width ($27~nm$) is limited by the channel-to-tall via space and Omni 3D channel width ($28~nm$) is defined by the gate extension; the same gate cut determines the location of a tall via and the gate edge of CFET and Omni 3D, respectively. Omni 3D NOR2 (e) without and (f) with IM present shortened intra-cell connection facilitating IM.}
\label{omni_3d}
\end{figure}

Conversely, silicon (Si) device architectures have continuously evolved and have been key in technology node scaling. Innovations such as nanosheets~\cite{nanosheet}, forksheets~\cite{forksheet}, and buried power rails (BPR)~\cite{bpr} have been proposed to enhance area efficiency, while back-side contacts (BSC)~\cite{bsc, bscon} and vertical-horizontal-vertical (VHV) structures~\cite{vhv} have been introduced to improve drive strength and intra-cell connectivity, respectively. Beyond such innovations, complementary FET (CFET) architectures (that stack pFETs and nFETs in 3D) provide another avenue for continued Si transistor scaling~\cite{intelcfet, tsmccfet, samcfet}; a few CFET alternatives have also been presented, however with relaxed design rules~\cite{ffet, dh_cfet}.

Recognizing the importance of device architecture, Omni 3D has been proposed as a dedicated solution for BEOL-compatible logic~\cite{omni}. In contrast to CFETs in Fig.~\ref{omni_3d} (a), Omni 3D's architecture (Fig.~\ref{omni_3d} (b)) is enabled by the following novelties: (1) One of the power rails is lifted above the upper FET to eliminate tall vias that limit the channel width in CFETs (Fig.~\ref{omni_3d} (c)). This enables Omni 3D $3$-track cell heights (Fig.~\ref{omni_3d} (d)) that maintain comparable drive strength of $4$-track CFETs. (2) Signal pins are defined on both the top and bottom sides as input (I) and output (Z) for double-side routing (See Fig.~\ref{omni_3d} (b)).
(3) Interleaved metal (IM) between nFET and pFET is introduced to provide extra intra-cell routing tracks. For example, Fig.~\ref{omni_3d} (e) and (f) shows how IM improves the routing for the parallel connection of nFETs in NOR2.
Prior work has shown how a combination of these advances allows Omni 3D to achieve $1.9\times$ energy-delay product (EDP) benefits in block-level designs (e.g., simple RISC-V cores) compared to CFET~\cite{omni}.

\begin{table}[!t]
    \centering
    \caption{Device design parameters at sub-$2~nm$ technology node}
    \begin{tabular}{|c|c|c|c|} \hline  
         Design parameters&  CFET&Omni 3D& Reference\\ \hline 
         Contacted gate pitch (nm)&  \multicolumn{2}{|c|}{42}& \cite{bsc, cfet} \\ \hline  
         Gate length (nm)&  \multicolumn{2}{|c|}{14, 15, 16, 17}& \cite{cnt7x} \\ \hline  
         Gate-to-S/D space (nm)&  \multicolumn{2}{|c|}{5, 7, 9}& \cite{cnt7x} \\ \hline  
         Gate cut (nm)&  \multicolumn{2}{|c|}{9}& \cite{bsc} \\ \hline  
         Gate extension (nm)&  \multicolumn{2}{|c|}{8.5}& \cite{bsc, cfet} \\ \hline  
         S/D extension (nm)&  \multicolumn{2}{|c|}{0}& \cite{cfet} \\ \hline  
         S/D-to-via space (nm)& 9& NA& \cite{3tcfet} \\ \hline  
         S/D-to-BPR space (nm)&  3&  NA& \cite{bsc} \\ \hline  
         M1 pitch/width (nm)&  \multicolumn{2}{|c|}{18/9}& \cite{bsc} \\ \hline  
         \# nanosheets&  \multicolumn{2}{|c|}{1, 2, 3, 4}& \cite{cnt7x} \\ \hline  
 $V_{dd}$ (V)& \multicolumn{2}{|c|}{0.45, 0.5, 0.55, 0.6, 0.65, 0.7}& \cite{cnt7x} \\ \hline 
    \end{tabular}
    \label{dtco_space}
\end{table}

Beyond the prior work on Omni 3D, we accomplish three key advances:
\begin{itemize}
    \item \textbf{\textit{Device model}}: While the prior work employed a predictive model from an existing Si CFET literature~\cite{cfet}, here we use a calibrated BEOL CNFET model. CNFETs are chosen as an example BEOL-compatible technology as: (1) CNFETs have large projected EDP benefits over Si transistors at advanced technology node~\cite{cnt7x}, (2) BEOL-integration of complementary CNFETs is already achieved within industrial Si fabs and foundries ~\cite{skywater}, and (3) complex BEOL circuits and systems have been demonstrated using CNFETs (a RISC-V core, the largest BEOL-compatible logic tapeout to date~\cite{cnt_riscv}).
    \item \textbf{\textit{Device architecture}}: Prior work showed a comparison of one Omni 3D option over several CFETs. However, in this work, we explore multiple Omni 3D variants by including/excluding IM and configuring pin access patterns to further optimize Omni 3D routing.
    \item \textbf{\textit{Physical design}}: The past work on Omni 3D evaluated a small benchmark (a design with $\leq 2K$ gates), due to the lack of a dedicated physical design flow (requiring manual customization of Omni 3D layouts). In contrast, we have now established a new Omni 3D physical design flow, integrated with commercial EDA tools supporting designs of various complexities (here, we show designs with $35K - 420K$ gates) through cell grouping and flipping algorithms, avoiding redundant routing and balancing metal usage on both sides.
\end{itemize}

The remainder of this paper consists as follows: We optimize device layout in Section~\ref{sec:dtco} and explore impact of IM and pin access patterns at standard cell level in Section~\ref{sec:variants}. Section~\ref{sec:pd} explains physical design challenges with existing commercial tools and their solutions. In Section~\ref{sec:exp}, Omni 3D is assessed and analyzed in comparison to CFET for three different logic cores. We summarize our work and discuss future opportunities in Section~\ref{sec:conclusion}.

\section{Design Technology Co-Optimization (DTCO)}
\label{sec:dtco}

\begin{figure}[!t]
\centering
\includegraphics[width=3.5in]{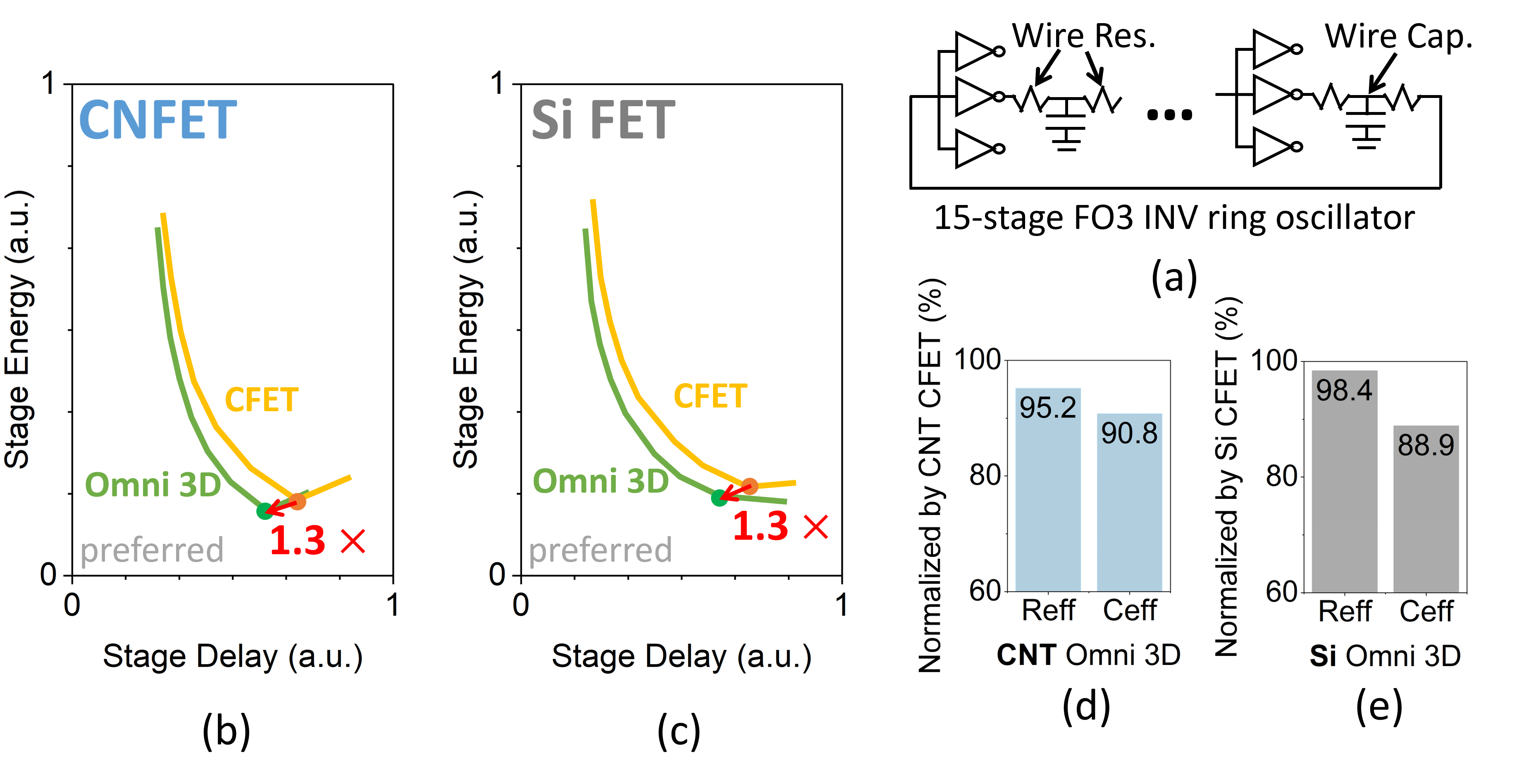}
\caption{(a) Benchmark circuits for DTCO: wire-loaded 15-stage FO3 INV RO. Energy vs. delay pareto curves of CFET and Omni 3D from our DTCO framework with (b) CNFETs and (c) Si FETs (results from~\cite{omni}). By coincidence, both channel materials achieve similar EDP benefits in Omni 3D ($\sim1.3\times$). However, the breakdown of $R_{eff}$ and $C_{eff}$ benefits of Omni 3D with (d) carbon nanotube (CNT) and (e) Si are different.}
\label{dtco_result}
\end{figure}

Device layouts vary with design parameters (e.g., gate length, gate-to-source/drain (S/D) space, and \# nanosheets), which impact EDP~\cite{cnt7x}. Our DTCO framework determines the best sets of design parameters for Omni 3D and CFET. We use a hardware-calibrated~\cite{vscnfet} and theoretically-refined~\cite{cnt7x} virtual-source CNFET model. A ring oscillator (RO) circuit with 15-stage minimum size INVs, fan-out 3 (FO3), and interconnect (Fig.~\ref{dtco_result} (a)) is employed as a benchmark~\cite{ro}. We explore a design space in Table~\ref{dtco_space} targeting a sub-$2~nm$ technology node. Cell parasitics of 3D layouts are extracted with GTS Cell Designer~\cite{gts_cd}. Before of every RO simulation, threshold voltages are re-targeted to set the leakage current of both nFET and pFET to $2~nA$/FET. Among $288$ combinations for each CFET and Omni 3D design, those with contact length less than $10~nm$ and those that cannot meet leakage current constraints by threshold adjustment are dropped from the set of feasible design points~\cite{cnt7x}.

We show the single stage RO energy vs. delay pareto curves for Omni 3D with CNFET channel in Fig.~\ref{dtco_result} (b). Minimum-EDP design points for CFET and Omni 3D are highlighted. Both design points have gate length of $14~nm$, gate-to-S/D space of $9~nm$, one nanosheet, and $V_{dd}$ of $0.45~V$. Omni 3D achieves $1.3\times$ EDP benefits over CFET. Energy and delay are respectively improved by $10.2\%$ and $15.6\%$.

Similar ($\sim1.3\times$) EDP benefits were reported with Si FETs~\cite{omni} as shown in Fig.~\ref{dtco_result} (c). However, the benefits stem from different reasons.
Energy ($=C_{eff}V_{dd}^2$) and delay ($=R_{eff}C_{eff}$) benefits in Omni 3D can be decomposed into improvements in effective capacitance ($C_{eff}$) and effective resistance ($R_{eff}$). $R_{eff}$ and $C_{eff}$ improvements for Omni 3D with CNFETs and Si FETs (compared to respective CFET architectures) are shown in Fig.~\ref{dtco_result} (d) and (e). The $1~nm$ widened channel in Omni 3D impacts the effective width more for CNFETs due to its thinner thickness ($1~nm$ vs. $5~nm$ of Si FET), further reducing $R_{eff}$. However, the benefits to $C_{eff}$ from track height reduction are diminished because CNFET's longer gate-to-S/D space ($9~nm$ vs. $5~nm$ of Si FET) lessens the corresponding capacitance fraction. Structural differences between Si FETs and CNFETs have minimal contribution to the EDP benefits.

\section{Omni 3D Variants}
\label{sec:variants}

We discuss two independent variations of the Omni 3D standard cell library: (1) IM inclusion/exclusion and (2) different configurations of input \& output pin within a cell.
 
\subsection{Impact of IM}

\begin{figure}[!t]
\centering
\includegraphics[width=3.4in]{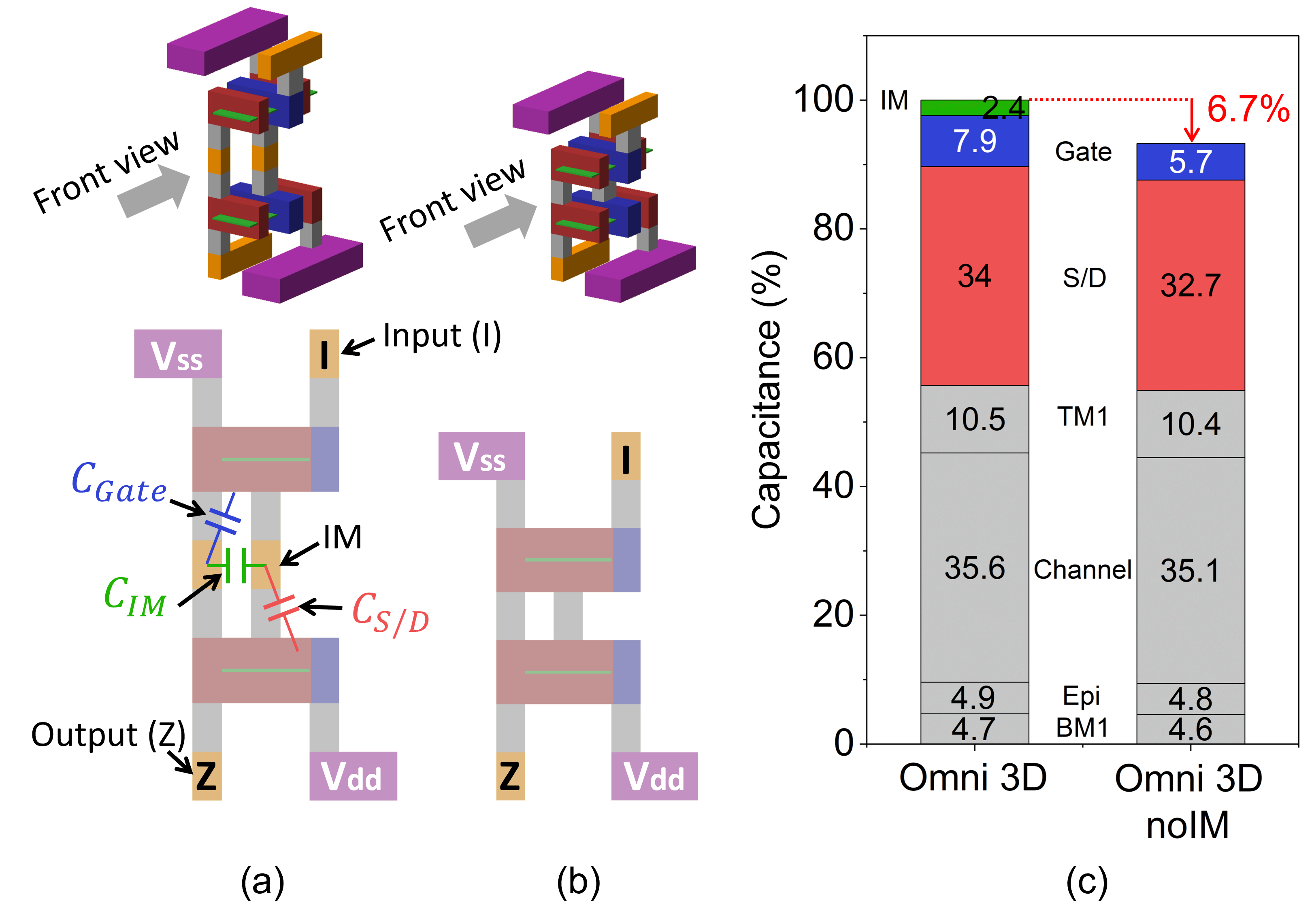}
\caption{Front view of (a) Omni 3D and (b) Omni 3D noIM INVs. (c) Their input capacitance breakdown. Capacitance of IM, gate, and S/D affected by IM elimination are color-coded. TM1: top-side metal 1 and BM1: bottom-side metal 1.}
\label{noIM}
\end{figure}

IM provides design flexibility with additional intra-cell routing tracks.
However, these routing tracks may introduce extra parasitic capacitance.
For example, INVs of Omni 3D and Omni 3D without IM (noIM) are illustrated in Fig.~\ref{noIM} (a) and (b), respectively; eliminating IM also removes the via below.
Fig.~\ref{noIM} (c) shows input capacitance reduction by $6.7\%$ in noIM. More specifically, IM to the other IM, gate, and S/D capacitance are shrunk. This savings is valid for cells (e.g., INV and BUF) which uses IM as only another via stack.

\begin{figure}[!t]
\centering
\includegraphics[width=3.0in]{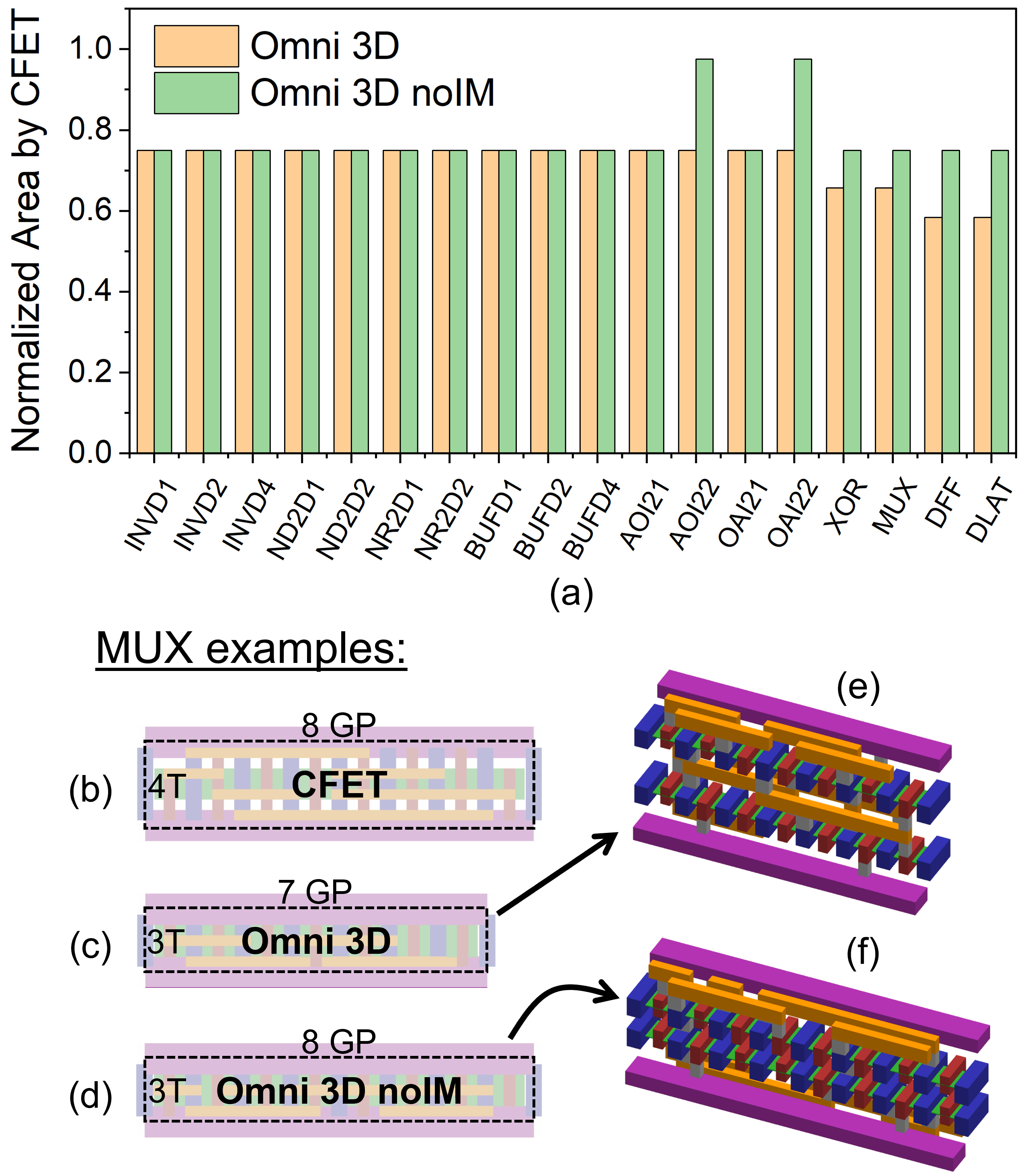}
\caption{(a) Area of Omni 3D and Omni 3D noIM normalized to corresponding CFETs for 20 basic cells. Area benefits exemplified with MUX top views: (b) CFET - $4$ tracks (T) $\times$ $8$ gate pitches (GP), (c) Omni 3D - $3$ T $\times$ $7$ GP, and (d) Omni 3D noIM - $3$ T $\times$ $8$ GP. MUX 3D illustrations of (e) Omni 3D and (f) Omni 3D noIM.}
\label{cell_area}
\end{figure}

However, the design impact of IM on complex cells is different from INV/BUFs. We implement $20$ standard cells for Omni 3D, Omni 3D noIM, and CFET. Fig.~\ref{cell_area} shows the area benefits of two Omni 3D variants over CFET and the impact of IM to the benefits. 
The majority of cells save $25\%$ area by cell height reduction, from $4$ tracks to $3$ tracks. Complex cells which demand heavy intra-cell routing (e.g., DFF, XOR, MUX) leverage extra three intra-cell routing tracks in the IM layer to further reduce cell area by shortening cell width in Omni 3D (orange in Fig.~\ref{cell_area} (a), See MUX examples in (c) vs. (b)). 
In contrast, noIM only achieves cell height reduction (green in Fig.~\ref{cell_area} (a), See (d) vs. (b)). This is because noIM has the same number of intra-cell routing metal tracks (two each on top and bottom) as CFET (four on top). 
IM usage for more compact Omni 3D MUX design is shown in Fig.~\ref{cell_area} (e) vs. (f).

AOI22 and OAI22 have notable area overhead in Omni 3D noIM.
One side of the Omni 3D noIM has only two routing tracks, which are fully occupied by the pins of such a many-input cell, leaving no room for intra-cell routing.
Cell width of noIM must therefore be extended to facilitate the necessary intra-cell routing on that side while Omni 3D achieves the routing on the IM layer and avoids cell width extension.

\subsection{Pin Access Pattern}

\begin{figure}[!t]
\centering
\includegraphics[width=3.0in]{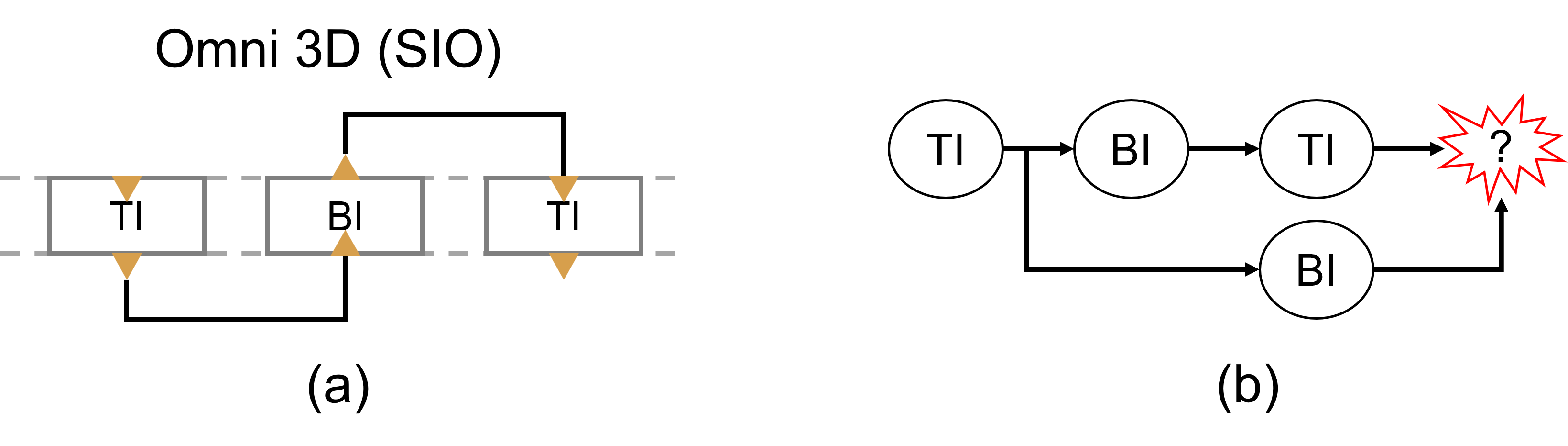}
\caption{(a) Side view of Omni 3D (SIO) routing example and (b) its routing sequential dependency.}
\label{route_limit}
\end{figure}

\begin{figure}[!t]
\centering
\includegraphics[width=3.4in]{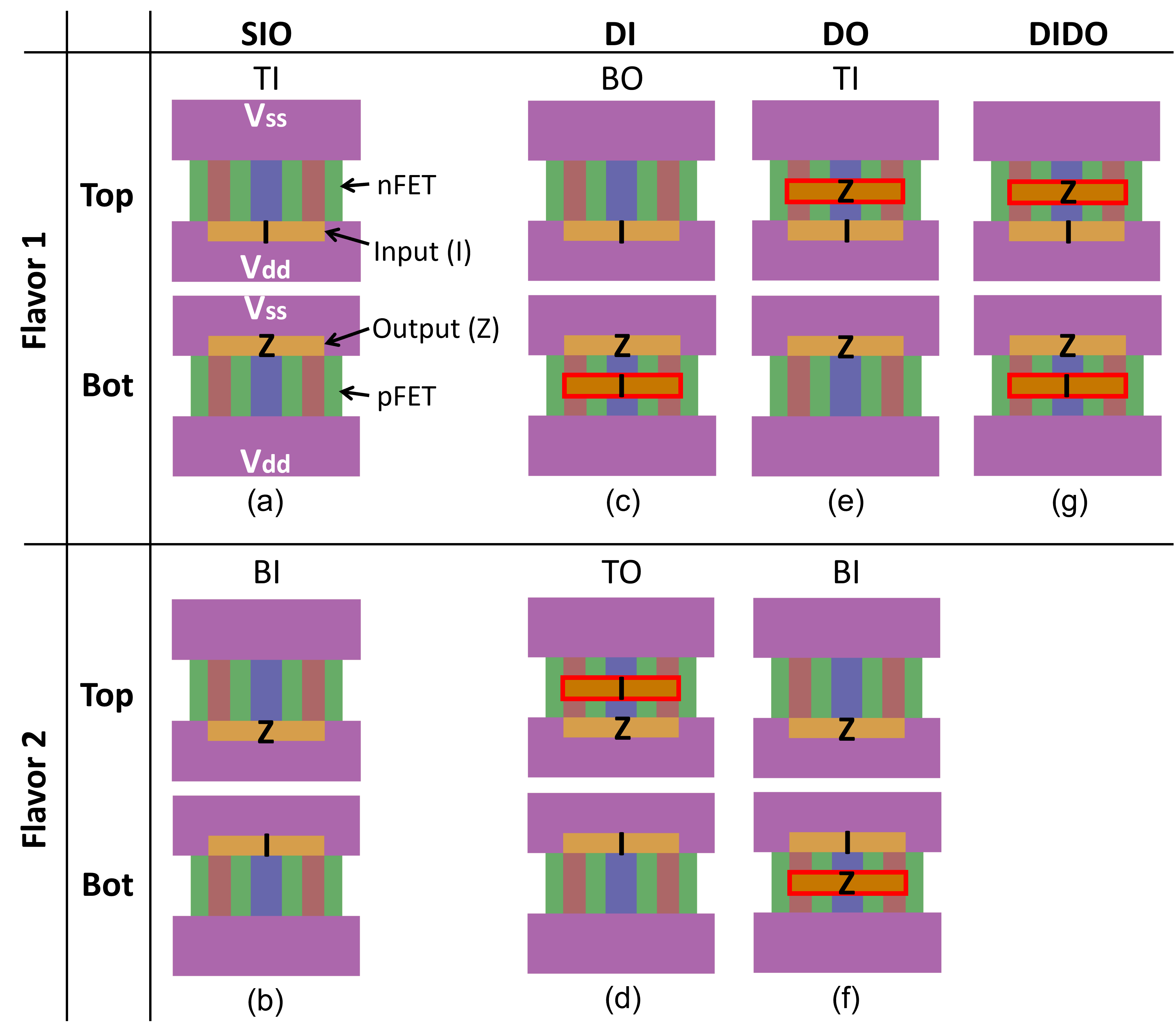}
\caption{Two SIO flavors of a Omni 3D INV: (a) TI and (b) BI. Omni 3D INV pin access variants from (a) and (b): adding an extra input to the bottom and top, respectively, DI of (c) BO and (d) TO; adding an extra output to the top and bottom, respectively, DO of (e) TI and (f) BI; adding extra inputs and outputs to the bottom and top, respectively, (g) DIDO}
\label{pa_variants}
\end{figure}

Input and output pin access configurations affect double-side signal routing.
Omni 3D with a single-side input and output (SIO) needs two flavors for routing: a top-in (TI) cell and a bottom-in (BI) cell. Fig.~\ref{route_limit} (a) depicts a chain of INV cells alternating between TI and BI to opimize routing. However, an arbitrary netlist can easily create a conflict due to multiple fan-in cells (in red) as exemplified in Fig.~\ref{route_limit} (b). Thus, such a sequential dependency restricts routing.

\begin{figure}[!t]
\centering
\includegraphics[width=3.0in]{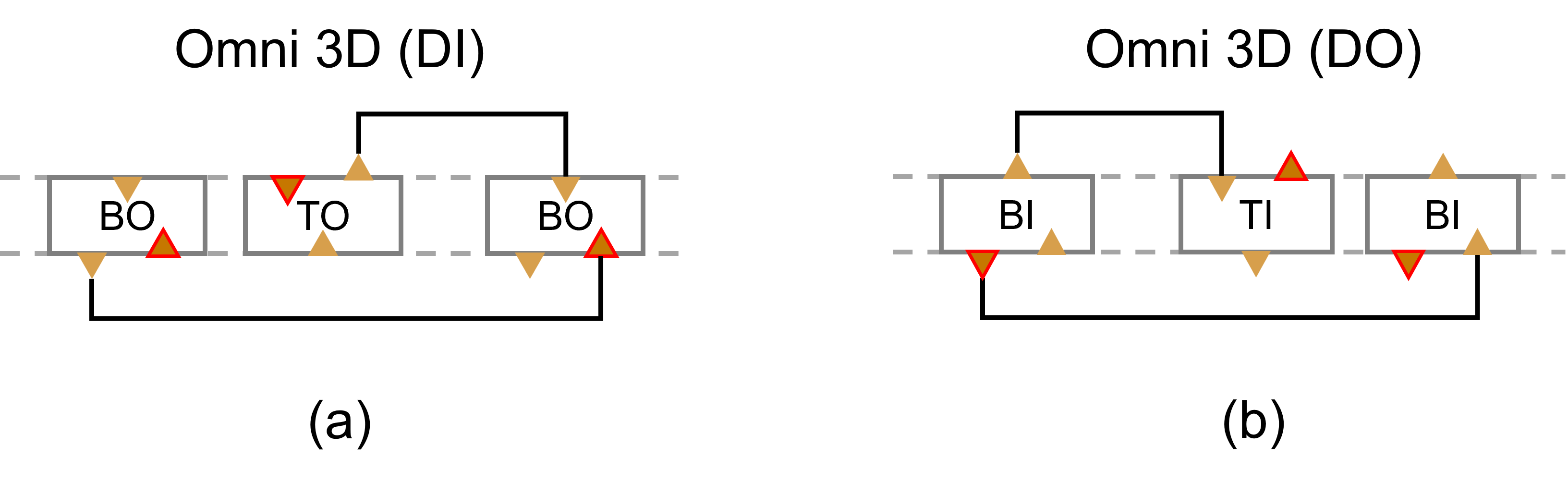}
\caption{Two Omni 3D variants' routing examples: (a) DI driven by any flavor of DI (b) DO driving any flavor of DO. Added input or output pins in red triangles are color-coded with Fig.~\ref{pa_variants}.}
\label{dido}
\end{figure}

Enabling variants of Omni 3D relieve this sequential dependency. One is a double-side input (DI) cell which features duplicated input pins on both sides, and another is double-side output (DO) cell that is characterized by duplicated output pins on both sides. By adding an input pin to the bottom of and an output pin to the top of Fig.~\ref{pa_variants} (a), respectively, bottom-out (BO) DI (Fig.~\ref{pa_variants} (c)) and TI DO (Fig.~\ref{pa_variants} (e)) are formed. From the other flavor of SIO, the corresponding top-out (TO) DI and BI DO are produced (See Fig.~\ref{pa_variants} (b), (d), and (f)).
Any DI cells can be routed by both BO and TO DI drivers as illustrated in Fig.~\ref{dido} (a), and any DO drivers can route both TI and BI DO cells that follow as presented in Fig.~\ref{dido} (b).
While double-side input and double-side output (DIDO) variants (Fig.~\ref{pa_variants} (g)) simplify the cell library, it results in excessively many pins on both sides.

\begin{figure}[!t]
\centering
\includegraphics[width=2.1in]{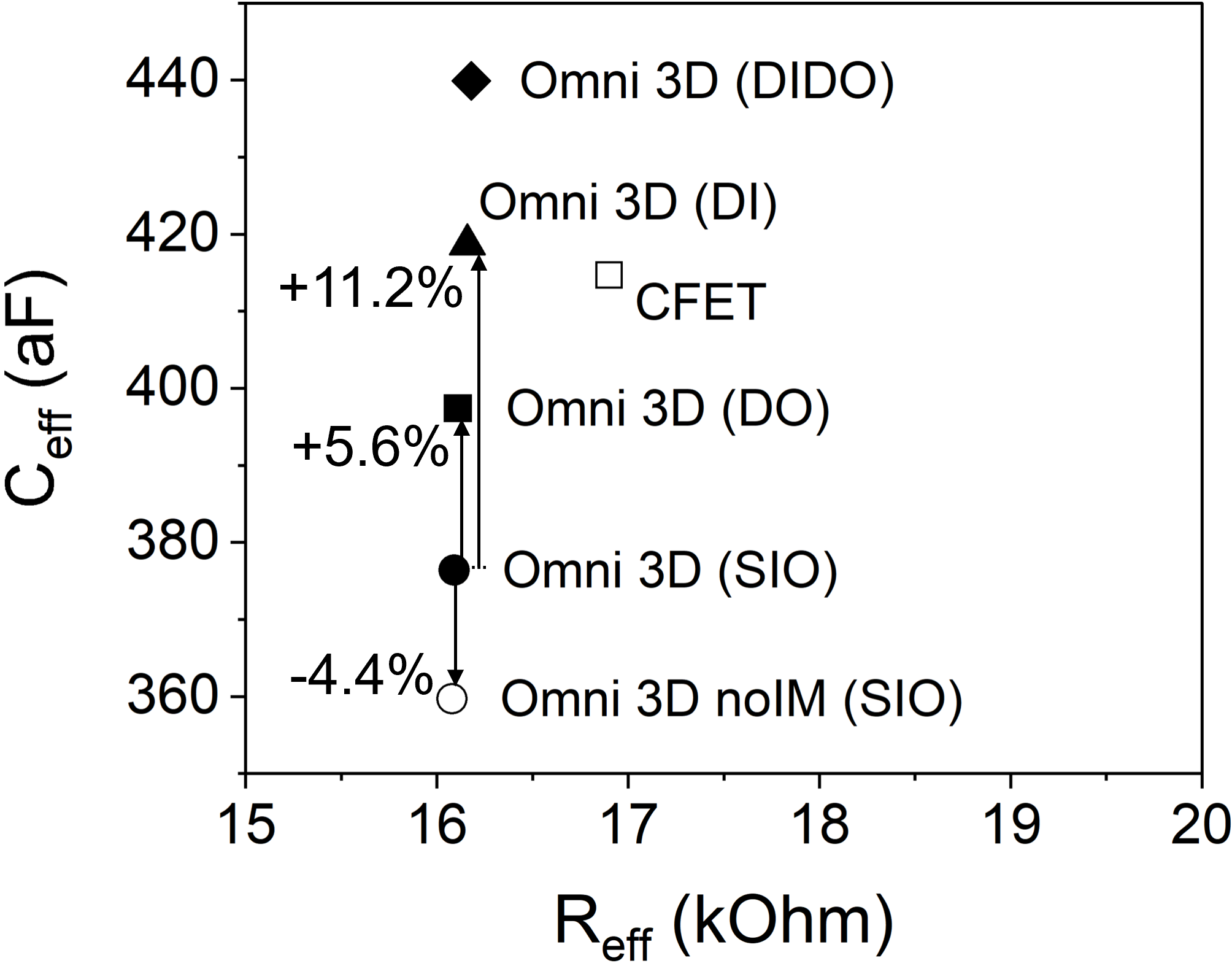}
\caption{$C_{eff}$  vs. $R_{eff}$ of minimum-EDP Omni 3D variants and CFET.}
\label{variants}
\end{figure}

\subsection{Discussion}

We evaluated all four variants of Omni 3D with the same DTCO framework in Section~\ref{sec:dtco}, and the minimum-EDP design points are shown in Fig.~\ref{variants}. All variants have similar $R_{eff}$s, $\sim95\%$ of CFET $R_{eff}$, because their channel widths are equal. Even though DO and DI both adds a pin to one side, the impact on $C_{eff}$ turns out to be different. While adding an extra output pin increases input and output capacitance by $4.2\%$ and $15.8\%$, respectively, adding an extra input pin raises input and output capacitance by $21.8\%$ and $5.8\%$, respectively. In terms of total cell parasitic capacitance, their difference is less than $3\%$ which is diluted by the interconnect capacitance of RO. However, DO and DI showed $5.6\%$ and $11.2\%$ $C_{eff}$ penalties, respectively, compared to SIO. This is because the input capacitance driven at the end of the interconnect has a greater impact on delay than the output capacitance driven in the beginning of the interconnect. Additionally, DIDO has $16.9\%$ more $C_{eff}$ as the sum of DI and DO increases $C_{eff}$ by $(\simeq 5.6\% + 11.2\%)$. NoIM  results in a $4.4\%$ $C_{eff}$ reduction. Its area benefits in complex cells, however, can only be assessed at the block level.

More importantly, DO always adds only one pin to SIO while DI, in the case of a multi-input cell, needs to place multiple pins which may demand extra design area to ensure intra-cell routing. Considering both $C_{eff}$ benefit and compact design, we choose DO as our optimal Omni 3D for block implementations. 

\section{Physical Design Enablement}
\label{sec:pd}

\subsection{PDK Preparation}
\label{subsec:setup}

\begin{figure}[!t]
\centering
\includegraphics[width=3.0in]{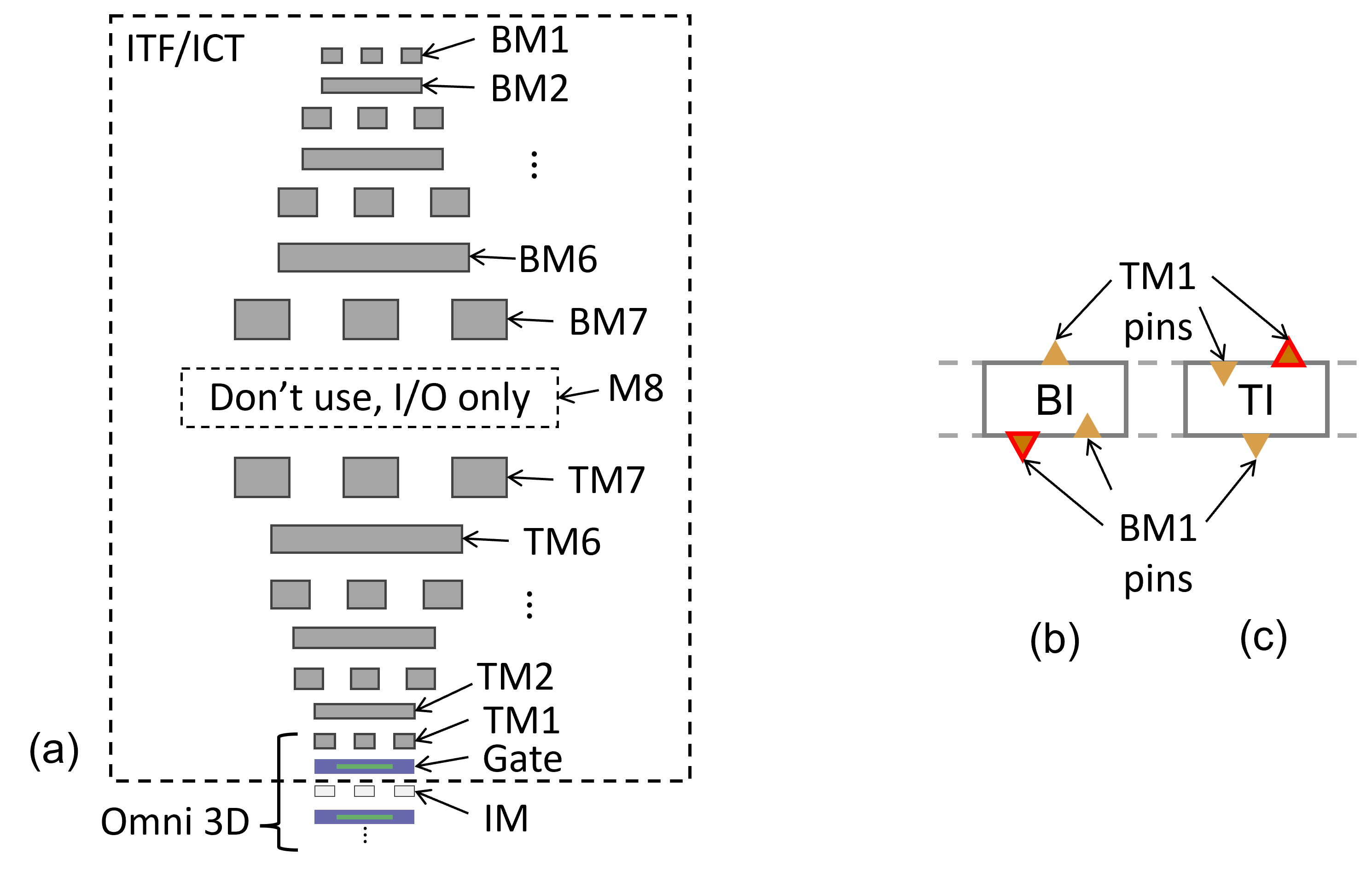}
\caption{Key changes in the PDK to enable double-side routing: (a) BEOL metal stack on top of Omni 3D in ITF/ICT; the focus is on BEOL metal parasitics, so the upper gate layer is only included in the ITF/ICT to follow convention. Pin layer definitions of (b) BI and (c) TI DO INVs in standard cell LEF file.}
\label{itf_lef}
\end{figure}

Two key changes in the process development kit (PDK) required to enable double-side routing with commercial EDA tools are: (1) An interconnect technology file (ITF/ICT) and (2) a standard cell layout exchange format (LEF) file.
Existing parasitic extraction tools which generate the interconnect RC database (QRC techfile/TLUPlus) only support metal stacks on top of the substrate.
Thus, we define bottom-side metal (BM) on top of top-side metal (TM) in reverse order as illustrated in Fig.~\ref{itf_lef} (a). One extra intermediate layer between TM and BM (i.e., M8) allows block I/O access on both sides but prohibits signal and clock routing to avoid a shortcut between TM7 to BM7. Such a shortcut would imply a vertical path penetrating the whole metal stack. We use predicted $2~nm$ technology node metal/via pitches, resistances, and capacitance~\cite{beol, irds}.

Fig.~\ref{itf_lef} (b) presents pin definitions of BI DO INV in the LEF file. An input pin and output pin on bottom are in BM1, and the other output pin on top is in TM1; vice versa for TI DO INV in Fig.~\ref{itf_lef} (c).
Pins with BM1 and TM1 are routed through BM2 $-$ BM7 and TM2 $-$ TM7, respectively.
Consistently, $V_{dd}$ and $V_{ss}$ power rails of each cell are defined on BM1 and TM1, respectively, in the LEF file.

Omni 3D has a split PDN, with $V_{dd}$ on the bottom side and $V_{ss}$ on the top side.
We constructed a mirrored PDN on both sides mimicking tight-pitch PDNs known to meet IR drop requirements~\cite{probe3, imecpdn}. This is equivalent to a conventional top-side PDN with densities of $6\%$ and $15\%$ for the lowest and highest routing metal layers, respectively.
The split PDN may create inductive loops in the power grid; optimizing the power grid for minimum inductance falls beyond the scope of our study.

\subsection{Efficient Double-Side Routing}
\label{subsec:bal}

\begin{figure}[!t]
\centering
\includegraphics[width=2.2in]{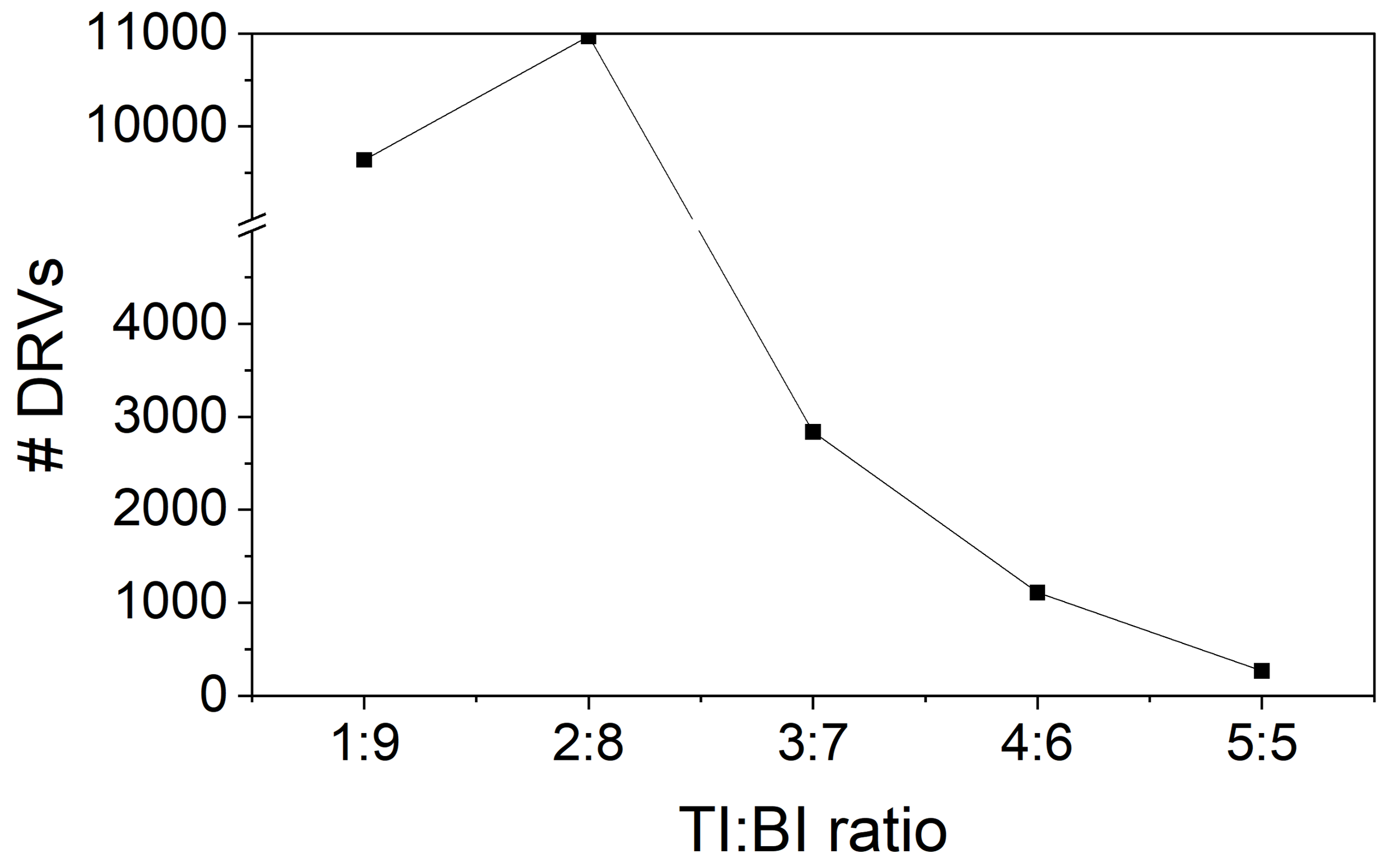}
\caption{\# DRVs along TI and BI cell ratio.}
\label{bal_motive}
\end{figure}

Traditional logic synthesis does not anticipate the Omni 3D physical design requirements in three ways. First, the netlist is synthesized to only considering power and delay, and the flavor (i.e., TI or BI) of a cell is arbitrarily determined. Taking the AES256 design as an example, the ratio of TI to BI is $1$:$8$. Using the resulting netlist would both diminish the value of Omni 3D as well as cause significant congestion. This issue can be addressed by balancing the TI and BI cell ratio after synthesis. Fig.~\ref{bal_motive} shows a decreasing trend of design rule violations (DRVs) as the TI:BI balance improves.

\begin{figure}[!t]
\centering
\includegraphics[width=2.6in]{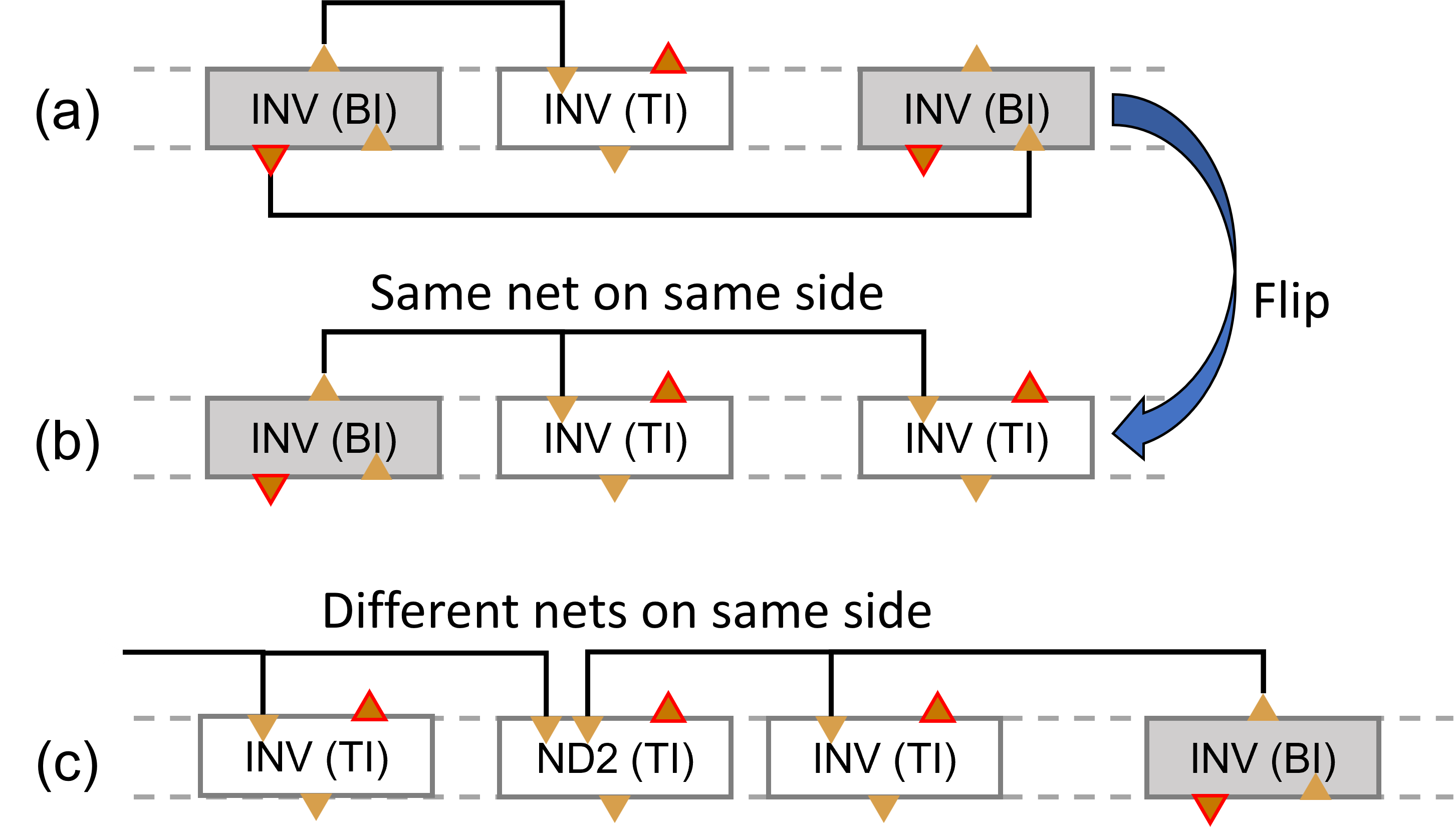}
\caption{(a) A logical net split to two physical nets on different sides. Cell flavor decision rules: (b) same net on same side to avoid redundant metal usage and (c) different nets sharing the cell inputs on same side.}
\label{cluster}
\end{figure}

Another aspect is that a logical net can be mapped to two different physical nets as shown in Fig.~\ref{cluster} (a). If an INV (BI) drives two different flavor INVs, a logical net is split into two sides to access the corresponding input pins. However, if either of the load cells is flipped to the other flavor as depicted in Fig.~\ref{cluster} (b), the two physical nets are united on one side eliminating redundant metal usage. 
Lastly, if two logical nets share a multi-input cell (e.g., ND2 in Fig.~\ref{cluster} (c)), they need to be assigned on the same side. This means that simply balancing the TI and BI cell ratio without considering these characteristics also induces unnecessary routing congestion. In AES256 with simple TI:BI cell count balancing, $6,575$ DRVs remained at the even ratio.

\begin{algorithm}[!t]
\small
  \caption{Cell Clustering Algorithm}
  \label{alg:clustering_algorithm}
  \begin{algorithmic}[1]  
  \REQUIRE $cellList$: list of cells
  \ENSURE $clusterList$: list of cell clusters
  \STATE $clusterList \leftarrow$ []
  \FORALL{$cell$ \textbf{in} $cellList$}
    \STATE $cluster \leftarrow$ []
    \STATE \textbf{call} \textsc{ClusterCell}($cell, cluster$)
    \STATE $clusterList$.append($cluster$)
  \ENDFOR

  \STATE

  \STATE \textbf{Function} \textsc{ClusterCell}($cell, cluster$)
    \IF{\textsc{IsCellClustered}($cell$)}
      \RETURN
    \ELSE
      \STATE $cluster$.append($cell$)
      \STATE $siblingCells \leftarrow$ \textsc{GetSiblingCells}($cell$)
      \FORALL{$sibling$ \textbf{in} $siblingCells$}
        \STATE \textbf{call} \textsc{ClusterCell}($sibling, cluster$)
      \ENDFOR
      \RETURN
    \ENDIF
  \STATE \textbf{End Function}

  \STATE

  \STATE \textbf{Function} \textsc{GetSiblingCells}($cell$)
    \STATE $siblingCells \leftarrow$ []
    \STATE $faninNets \leftarrow$ \textsc{GetFaninNets}($cell$)
    \FORALL{$net$ \textbf{in} $faninNets$}
      \STATE $siblingCells$.append(\textsc{GetFanoutCells}($net$))
    \ENDFOR
    \RETURN $siblingCells$
  \STATE \textbf{End Function}

  \end{algorithmic}
\end{algorithm}

To accommodate the Omni 3D physical design requirements for efficient double-side routing, we modify a netlist after logic synthesis by following Algo.~\ref{alg:clustering_algorithm}.
We first identify nets that need to be on the same side and accordingly cluster the relevant cells to be the same flavor. Depth-first search is used to find the cells to cluster. In detail, we start with putting a cell in $cellList$ into an empty cluster. In the function $getSiblingCells$, we obtain the cell's fan-in nets which must be on the same side, and then list out the sibling cells which are the fan-out cells of those nets. A sibling cell is added to the cluster by a recursive call of the function $clusterCell$, increasing the search depth. Depth drilling ends when no sibling cells are found or the cell of interest is already clustered. 

The clusters are then assigned to either a TI or BI flavor, targeting an even ratio of TI to BI. For simplicity, we sort the clusters in descending order by size and assign them alternatively to TI and BI flavors until the number of TI cells exceeds half of the total cell count. Large clusters are thus distributed evenly, with any remaining clusters assigned to BI.

\subsection{Design Implementation Flow}

We use the foundation flow~\cite{mflowgen} with three extra steps. (1) The cell clustering and assignment in Section~\ref{subsec:bal} are performed after synthesis, and the updated netlists are loaded by the placer, with the placer using cell clusters for its seed placement.
(2) Existing clock tree synthesis (CTS) inserts arbitrary flavors of buffers only considering delay and power, which leads to the clock tree using the prohibited layer (M8). After CTS, therefore, we flip the flavors of clock buffers whose fan-in nets employ M8.
(3) Upon completion of detailed routing, we also flip the flavors of buffers in data paths to address potential detoured routes --- nets that violate maximum transition time --- by the arbitrary-flavor buffer insertion.

\section{Assessment}
\label{sec:exp}

\begin{table*}[!t]
\centering
\caption{Summary of three benchmark designs' EDP and area benefits with top and bottom metal stack configurations.}
\begin{tabular}{|cc|ccc|ccr|ccr|}
\hline
\multicolumn{2}{|c|}{}                                             & \multicolumn{3}{c|}{CFET}                                                                     & \multicolumn{3}{c|}{Omni 3D}                                                                                                                                              & \multicolumn{3}{c|}{Omni 3D noIM}                                                                                                                                         \\ \hline
\multicolumn{2}{|c|}{Design}                                       & \multicolumn{1}{c|}{Rocket}       & \multicolumn{1}{c|}{LDPC}    & AES                        & \multicolumn{1}{c|}{Rocket}                                                         & \multicolumn{1}{c|}{LDPC}                    & \multicolumn{1}{c|}{AES}             & \multicolumn{1}{c|}{Rocket}                                                         & \multicolumn{1}{c|}{LDPC}                    & \multicolumn{1}{c|}{AES}             \\ \hline
\multicolumn{2}{|c|}{EDP}                                          & \multicolumn{3}{r|}{\multirow{4}{*}{$1.0\times$}}                                             & \multicolumn{1}{r|}{$2.3\times$}                                                    & \multicolumn{1}{r|}{$1.87\times$}             & $1.8\times$                          & \multicolumn{1}{r|}{$2.3\times$}                                                    & \multicolumn{1}{r|}{$1.75\times$}             & $1.8\times$                          \\ \cline{1-2} \cline{6-11} 
\multicolumn{2}{|c|}{Energy}                                       & \multicolumn{3}{r|}{}                                                                         & \multicolumn{1}{r|}{$1.2\times$}                                                    & \multicolumn{1}{r|}{$1.3\times$}             & $1.2\times$                          & \multicolumn{1}{r|}{$1.2\times$}                                                    & \multicolumn{1}{r|}{$1.3\times$}             & $1.2\times$                          \\ \cline{1-2} \cline{6-11} 
\multicolumn{2}{|c|}{Delay}                                        & \multicolumn{3}{r|}{}                                                                         & \multicolumn{1}{r|}{$1.9\times$}                                                    & \multicolumn{1}{r|}{$1.4\times$}             & $1.5\times$                          & \multicolumn{1}{r|}{$1.8\times$}                                                    & \multicolumn{1}{r|}{$1.4\times$}             & $1.5\times$                          \\ \cline{1-2} \cline{6-11} 
\multicolumn{2}{|c|}{Core Area}                                         & \multicolumn{3}{r|}{}                                                                         & \multicolumn{1}{r|}{$1.4\times$}                                                    & \multicolumn{1}{r|}{$1.6\times$}             & $1.4\times$                          & \multicolumn{1}{r|}{$1.2\times$}                                                    & \multicolumn{1}{r|}{$1.3\times$}             & $1.3\times$                          \\ \hline
\multicolumn{2}{|c|}{\# Cells (K)}                                 & \multicolumn{1}{r|}{35}           & \multicolumn{1}{r|}{56}      & \multicolumn{1}{r|}{419}   & \multicolumn{1}{r|}{37}                                                             & \multicolumn{1}{r|}{50}                      & 413                                  & \multicolumn{1}{r|}{38}                                                             & \multicolumn{1}{r|}{56}                      & 416                                  \\ \hline
\multicolumn{1}{|c|}{\multirow{2}{*}{Top metal}}    & Signal/Clock & \multicolumn{1}{c|}{TM2 - TM5}    & \multicolumn{2}{c|}{TM2 - TM7}                            & \multicolumn{1}{c|}{TM2 - TM5}                                                      & \multicolumn{2}{c|}{TM2 - TM7}                                                      & \multicolumn{1}{c|}{TM2 - TM5}                                                      & \multicolumn{2}{c|}{TM2 - TM7}                                                      \\ \cline{2-11} 
\multicolumn{1}{|c|}{}                              & Power        & \multicolumn{1}{c|}{-}            & \multicolumn{2}{c|}{-}                                    & \multicolumn{1}{c|}{\begin{tabular}[c]{@{}c@{}}TM2 - TM5\\ ($V_{ss}$)\end{tabular}} & \multicolumn{2}{c|}{\begin{tabular}[c]{@{}c@{}}TM2 - TM7\\ ($V_{ss}$)\end{tabular}} & \multicolumn{1}{c|}{\begin{tabular}[c]{@{}c@{}}TM2 - TM5\\ ($V_{ss}$)\end{tabular}} & \multicolumn{2}{c|}{\begin{tabular}[c]{@{}c@{}}TM2 - TM7\\ ($V_{ss}$)\end{tabular}} \\ \hline
\multicolumn{1}{|c|}{\multirow{2}{*}{Bottom metal}} & Signal/Clock & \multicolumn{1}{c|}{-}            & \multicolumn{2}{c|}{-}                                    & \multicolumn{1}{c|}{BM2 - BM5}                                                      & \multicolumn{2}{c|}{BM2 - BM7}                                                      & \multicolumn{1}{c|}{BM2 - BM5}                                                      & \multicolumn{2}{c|}{BM2 - BM7}                                                      \\ \cline{2-11} 
\multicolumn{1}{|c|}{}                              & Power        & \multicolumn{3}{c|}{\begin{tabular}[c]{@{}c@{}}BM2 - BM5\\ ($V_{dd}$, $V_{ss}$)\end{tabular}} & \multicolumn{1}{c|}{\begin{tabular}[c]{@{}c@{}}BM2 - BM5\\ ($V_{dd}$)\end{tabular}} & \multicolumn{2}{c|}{\begin{tabular}[c]{@{}c@{}}BM2 - BM7\\ ($V_{dd}$)\end{tabular}} & \multicolumn{1}{c|}{\begin{tabular}[c]{@{}c@{}}BM2 - BM5\\ ($V_{dd}$)\end{tabular}} & \multicolumn{2}{c|}{\begin{tabular}[c]{@{}c@{}}BM2 - BM7\\ ($V_{dd}$)\end{tabular}} \\ \hline
\end{tabular}
    \label{final_data}
\end{table*}

We implemented three open-source designs, a processor (Rocket)~\cite{rocket}, ECC core (LDPC), and crypto core (AES256)~\cite{opencore} with CFETs, Omni 3D, and noIM libraries; CFETs use BSPDNs. Each design was implemented and simulated with various target clock periods to obtain the minimum-EDP design point; sweep ranges were $300~ps - 100~ps$, $900~ps - 500~ps$, $200~ps - 60~ps$ with $20~ps$ interval for Rocket, LDPC, and AES256, respectively. For technology assessment purposes, we take the average slack and clock skew of top $100$ critical paths rather than just the top critical path to avoid oultier distortions ~\cite{probe3}. An achieved delay was calculated by subtracting the slack from the targeted clock period. Designs with \# DRVs $\leq 300$, slack $\geq -50~ps$, and clock skew $\leq 10~ps$ are considered as valid implementations. Setting up area-hungry design specifications, final cell densities of Rocket, LDPC, and AES256 spanned $83\% - 90\%$, $53\% - 63\%$, and $77\% - 89\%$, respectively.

\subsection{Results}

\begin{figure}[!t]
\centering
\includegraphics[width=3.4in]{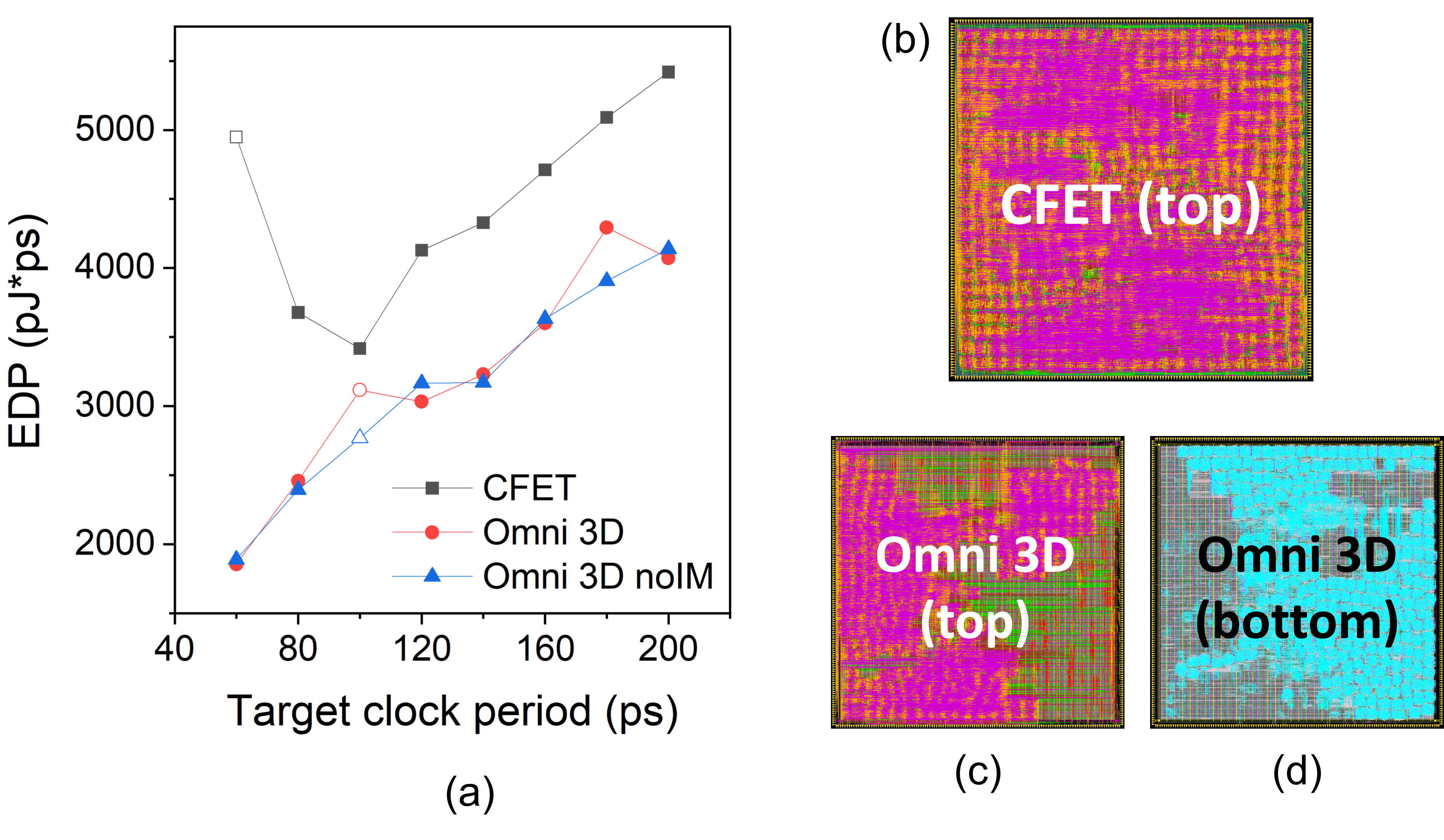}
\caption{AES256 physical design results: (a) EDP across target clock period, post-PnR layout of (b) CFET top, (c) Omni 3D top, and (d) Omni 3D bottom side.}
\label{aes}
\end{figure}

In a relaxed target clock period region, as the target decreases, design achieves lower delay without significant impact on energy consumption. Therefore, the EDP trends of AES256 using CFETs and Omni 3D in Fig.~\ref{aes} (a) are linear until $100~ps$; empty symbols are invalid designs but are included to show the trend. The EDP gap between the two libraries is similar to one shown in RO ($1.3\times$ EDP benefits in Omni 3D). However, after $100~ps$, CFETs started to include substantially more drivers to attain the target clock period and routing accordingly became more convoluted. However, Omni 3D  still had enough routing capabilities. This difference allows delay benefits in Omni 3D physical design over RO. In the comparison of minimum-EDP designs, Omni 3D earned $1.8\times$ EDP and $1.4\times$ area benefits over CFETs simultaneously. Post-PnR layouts of AES256 are in Fig.~\ref{aes} (b)$-$(d). An even spatial division for the top and bottom side routing is achieved for Omni 3D.

The EDP and area benefits for the all three designs compared to CFETs are summarized in Table~\ref{final_data}. Rocket has a small gate count ($\approx 35K$) which can be routed up to TM5 in CFETs, so Omni 3D used up to TM5 and down to BM5. EDP and area benefits of Omni 3D are, respectively, $2.0\times$ and $1.5\times$ on average. Energy benefits are capped by $1.3\times$ while delay benefits vary depending on the congestion level of designs. noIM achieved the same EDP benefits on average while its area was saved by only $1.3\times$ due to the cell-level area penalties appeared in Fig.~\ref{cell_area} (a).

\subsection{Analysis and Observations}

\begin{figure}[!t]
\centering
\includegraphics[width=3.4in]{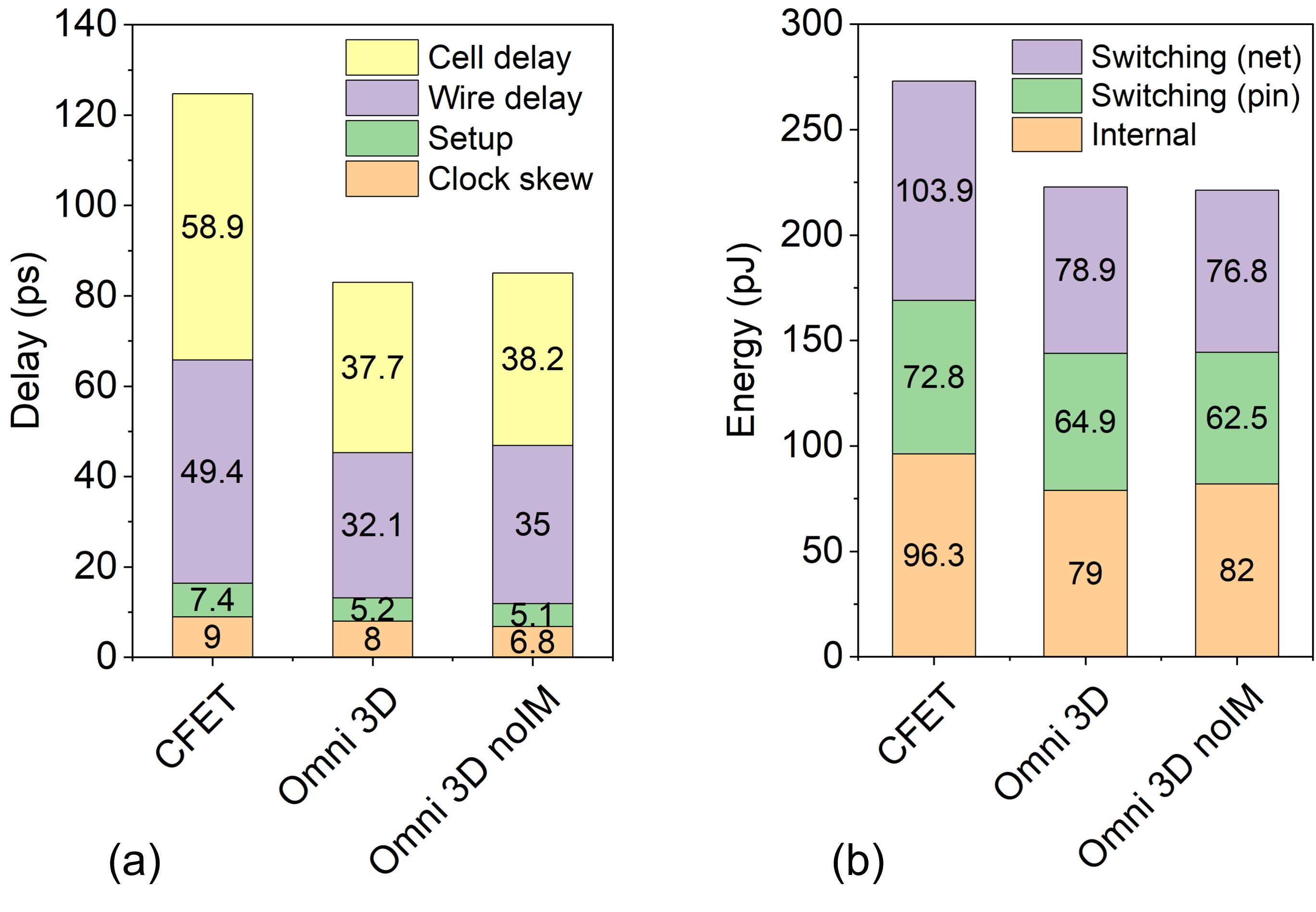}
\caption{AES256 (a) delay and (b) energy breakdown.}
\label{breakdown}
\end{figure}

We analyze the delay, energy, and area benefits in detail taking AES256 as an example, comparing Omni 3D to CFET and then to noIM.

\textbf{\textit {Delay:}} Fig.~\ref{breakdown} (a) shows the delay breakdown of the top $100$ critical paths of each library. The achieved delay is composed of cell delay, wire delay, setup time of the flip-flop, and clock skew. Before comparison, one notable observation for CFETs is that the wire delay, constituting $39.6\%$ of the total delay, approached that of the cell delay ($47.2\%$). This is attributed to the significant cell resistance improvement of using CNFETs.
By introducing Omni 3D, routing congestion was relaxed, hence both wire and driver delays were reduced  by $35.0\%$ and $36.0\%$, respectively. noIM showed negligible difference in cell delay from Omni 3D and a $9.0\%$ penalty in wire delay, due to core area increase.

\begin{figure}[!t]
\centering
\includegraphics[width=2.3in]{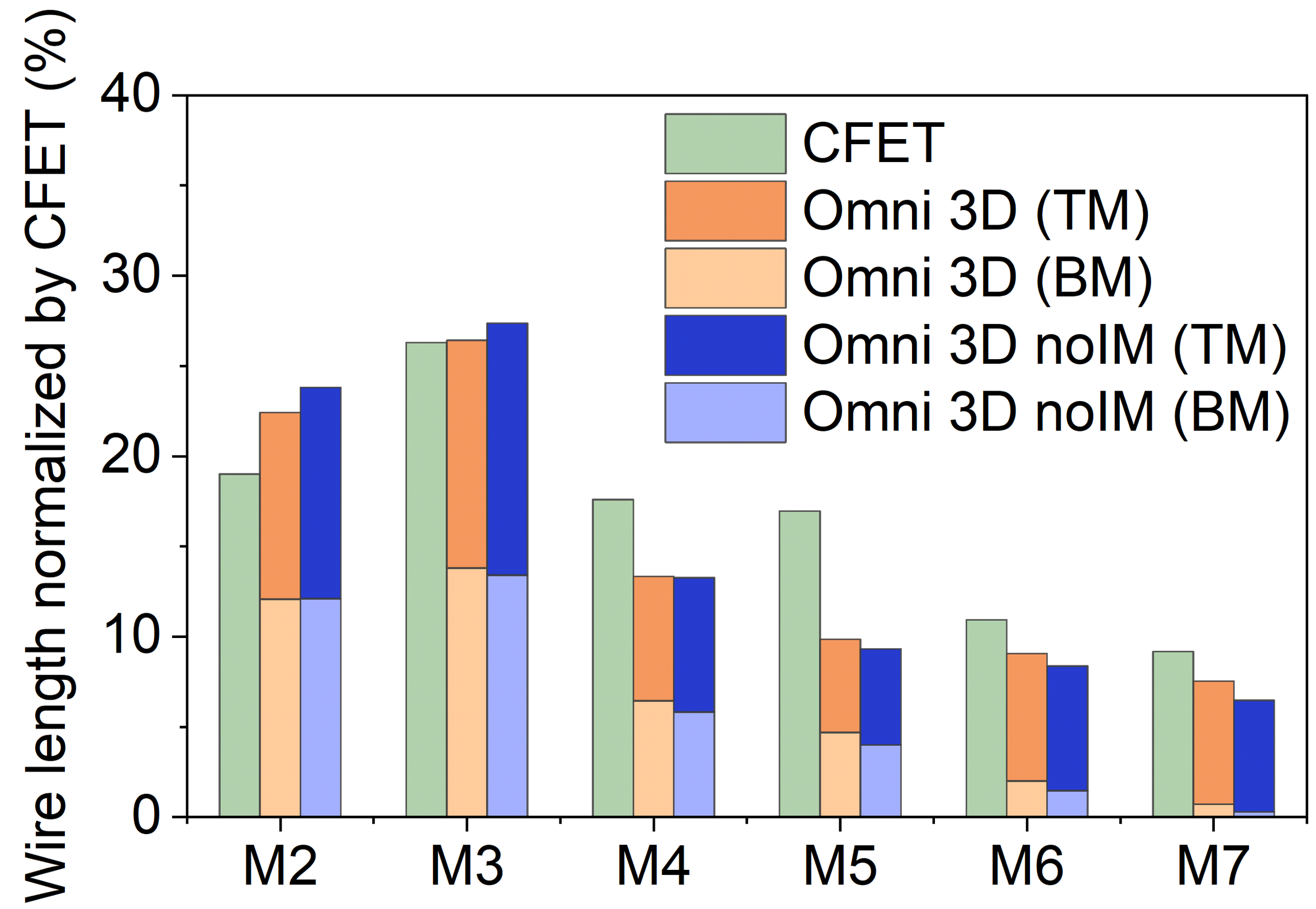}
\caption{AES256 wire length distribution across metal layers. Omni 3D M2 indicates the sum of TM2 and BM2.}
\label{wire_dist}
\end{figure}

\textbf{\textit {Energy:}} Energy breakdown is in Fig.~\ref{breakdown} (b). With a default logic switching rate, leakage energy is about $1\%$; dynamic energy is decomposed into switching energies and internal energies. Internal energy is estimated from internal power, and switching energy of pins and nets are estimated from switching power in proportion to their capacitances. While cell count is comparable ($1.4\%$ lower in Omni 3D compared to CFET), cell energy --- sum of pin switching and internal energy --- is lower by $14.9\%$ due to the compact cell design of Omni 3D. More importantly, net switching energy decreased by $24.6\%$ in Omni 3D even though the total wire length is only $11.4\%$ shorter. We attribute the extra net switching energy savings to changes in the wire length distributions across layers (see Fig.~\ref{wire_dist}). M2 of Omni 3D in total is more used than in CFET, but TM2 and BM2 are physically located on two different layers. Thus, metal density of Omni 3D in each layer, which contributes to both ground and coupling capacitance, is reduced. 

We observed that intensive use of lower metal layers (M2 and M3) in Omni 3D reduces upper layer metal (M4 and M5) utilization in Fig.~\ref{wire_dist}. Considering the $1.4\times$ area of CFET over Omni 3D, and that both M2 and M3 are similar lengths indicates that Omni 3D relies more on lower metal layers; in other words, CFET is congested in M2 and M3, so some paths are detoured through M4 and M5. noIM has less than $0.1\%$ difference to Omni 3D in total wire length, and its distribution is only slightly different.

\begin{figure}[!t]
\centering
\includegraphics[width=3.2in]{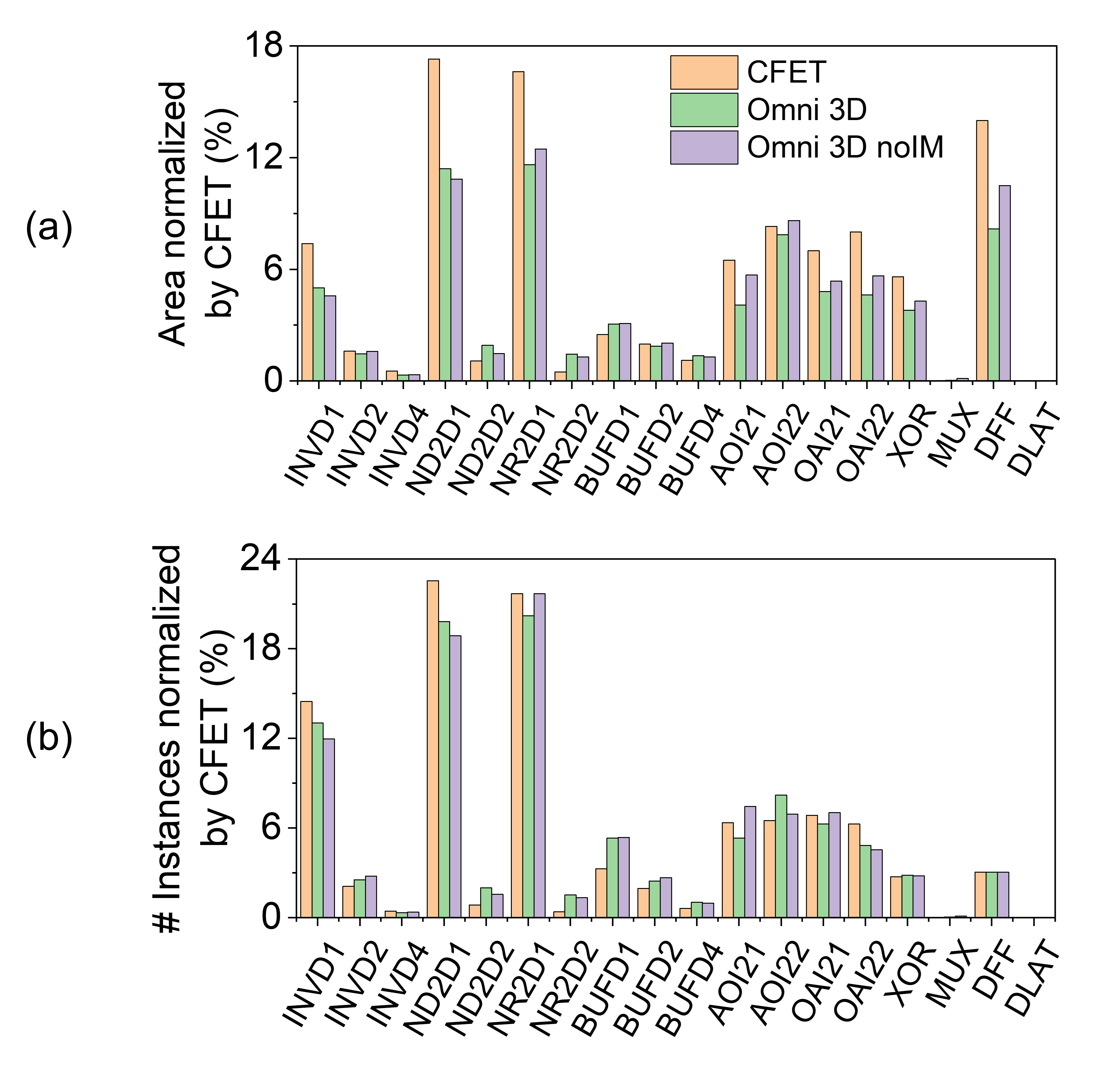}
\caption{Individual cell (a) area and (b) counts in AES256 normalized by CFET total area and cell counts, respectively.}
\label{area_analysis}
\end{figure}

\textbf{\textit {Area:}} To understand the $1.4\times$ area benefits in Omni 3D over CFET, area occupied by each cell in AES256 is normalized by the CFET AES256 total area in Fig~\ref{area_analysis} (a). CFET bars indicate the area contributions of individual cell. INVD1, ND2D1, NR2D1, and DFF have significant area contributions. Accordingly, area benefits of Omni 3D mainly comes from these cells. Note that while INV, ND, and NR provide only $1.3\times$ area benefit, $1.8\times$ area benefit of DFF is reflected in its large area savings contribution ($14.0\%$). Some other cells (e.g., BUFD1) occupy similar or more area compared to CFET because their cell count increases as presented in Fig.~\ref{area_analysis} (b). 

Fig.~\ref{area_analysis} (a) shows that DFF area benefits in Omni 3D are decreased in the noIM case vs. IM (due to additional routing flexibility in IM). Another notable impact of IM is that AOI and OAI distributions shift. Due to area-inefficient design of AOI22 and OAI22, their counts are lowered in Omni 3D noIM desgin points compared to Omni 3D designs, and instead, the counts of AOI21 and OAI21 increased to synthesize the same functions. Due to cell-level area increase in a few cells (e.g., DFF, AOI22, and OAI22), AES256 area benefit of noIM was cut to $1.3\times$.

\section{Conclusion}
\label{sec:conclusion}
In this work, we investigate Omni 3D device architecture with a BEOL-compatible channel and physical design for efficient double-side routing. Omni 3D (DO) with IM showed the highest EDP ($2.0\times$) and area ($1.5\times$) benefits compared to CFETs with BSPDNs. noIM, with a lesser area benefit of $1.3\times$ can serve as an alternative option if IM is not preferred for device stability, fabrication, or cost issues. 

Future work should address the EDA challenges of multi-tier logic for 3D systems and explore their design trade-offs systematically.
Critical path-aware cell clustering and side assignment further improve delay. A remaining metal usage imbalance between sides (Fig.~\ref{wire_dist} M6 and M7), due to naïve metal layer assignments, warrants further investigation. Additionally, multi-tier Omni 3D, with its full tier-to-tier routing capabilities needs to be evaluated against traditional face-to-face or face-to-back 3D logic that relies on single-side routing and TSVs.

\section*{Acknowledgment}
We thank SystemX Alliance for the support on this research.

\bibliographystyle{IEEEtran}
\bibliography{reference}

\begin{thebibliography}{10}
\providecommand{\url}[1]{#1}
\csname url@samestyle\endcsname
\providecommand{\newblock}{\relax}
\providecommand{\bibinfo}[2]{#2}
\providecommand{\BIBentrySTDinterwordspacing}{\spaceskip=0pt\relax}
\providecommand{\BIBentryALTinterwordstretchfactor}{4}
\providecommand{\BIBentryALTinterwordspacing}{\spaceskip=\fontdimen2\font plus
\BIBentryALTinterwordstretchfactor\fontdimen3\font minus \fontdimen4\font\relax}
\providecommand{\BIBforeignlanguage}[2]{{%
\expandafter\ifx\csname l@#1\endcsname\relax
\typeout{** WARNING: IEEEtran.bst: No hyphenation pattern has been}%
\typeout{** loaded for the language `#1'. Using the pattern for}%
\typeout{** the default language instead.}%
\else
\language=\csname l@#1\endcsname
\fi
#2}}
\providecommand{\BIBdecl}{\relax}
\BIBdecl

\bibitem{omni}
T.~Srimani \emph{et~al.}, ``{N3XT 3D Technology Foundations and Their Lab-to-Fab: Omni 3D Logic, Logic+ Memory Ultra-Dense 3D, 3D Thermal Scaffolding},'' in \emph{IEEE IEDM}, 2023, pp. 1--4.

\bibitem{n3xt}
M.~M.~S. Aly \emph{et~al.}, ``{The N3XT approach to energy-efficient abundant-data computing},'' \emph{Proceedings of the IEEE}, vol. 107, no.~1, pp. 19--48, 2018.

\bibitem{lowd}
S.-K. Su \emph{et~al.}, ``{Perspective on low-dimensional channel materials for extremely scaled CMOS},'' in \emph{IEEE Symp. on VLSI Tech.}, 2022, pp. 403--404.

\bibitem{2d}
Y.-Y. Chung \emph{et~al.}, ``{First Demonstration of GAA Monolayer-MoS 2 Nanosheet nFET with 410$\mu$A $\mu$ m ID 1V VD at 40nm gate length},'' in \emph{IEEE IEDM}, 2022, pp. 34--5.

\bibitem{cnt}
G.~Pitner \emph{et~al.}, ``{Building high performance transistors on carbon nanotube channel},'' in \emph{IEEE Symp. on VLSI Tech.}, 2023, pp. 1--2.

\bibitem{back}
------, ``{Sub-0.5 nm interfacial dielectric enables superior electrostatics: 65 mV/dec top-gated carbon nanotube FETs at 15 nm gate length},'' in \emph{IEEE IEDM}, 2020, pp. 3--5.

\bibitem{top}
S.~Li \emph{et~al.}, ``{High-performance and low parasitic capacitance CNT MOSFET: 1.2 mA/$\mu$m at V DS of 0.75 V by self-aligned doping in sub-20 nm spacer},'' in \emph{IEEE IEDM}, 2023, pp. 1--4.

\bibitem{gaa}
N.~Safron \emph{et~al.}, ``{High Performance Transistor of Aligned Carbon Nanotubes in a Nanosheet Structure},'' in \emph{IEEE Symp. on VLSI Tech.}, 2024, pp. 1--2.

\bibitem{barrier}
C.~Gilardi \emph{et~al.}, ``{Barrier Booster for Remote Extension Doping and its DTCO for 1D \& 2D FETs},'' in \emph{IEEE IEDM}, 2023, pp. 1--4.

\bibitem{nanosheet}
N.~Loubet \emph{et~al.}, ``{Stacked nanosheet gate-all-around transistor to enable scaling beyond FinFET},'' in \emph{IEEE Symp. on VLSI Tech.}, 2017, pp. T230--T231.

\bibitem{forksheet}
P.~Weckx \emph{et~al.}, ``{Novel forksheet device architecture as ultimate logic scaling device towards 2nm},'' in \emph{IEEE IEDM}, 2019, pp. 36--5.

\bibitem{bpr}
A.~Gupta \emph{et~al.}, ``{Buried power rail integration with FinFETs for ultimate CMOS scaling},'' \emph{IEEE Trans. on Electron Devices}, vol.~67, no.~12, pp. 5349--5354, 2020.

\bibitem{bsc}
S.~Yang \emph{et~al.}, ``{PPA and scaling potential of backside power options in N2 and A14 nanosheet technology},'' in \emph{IEEE Symp. on VLSI Tech.}, 2023, pp. 1--2.

\bibitem{bscon}
M.~Kobrinsky \emph{et~al.}, ``{Novel cell architectures with back-side transistor contacts for scaling and performance},'' in \emph{IEEE Symp. on VLSI Tech.}, 2023, pp. 1--2.

\bibitem{vhv}
V.~Vega-Gonzalez \emph{et~al.}, ``{Semi-damascene integration of a 2-layer MOL VHV scaling booster to enable 4-track standard cells},'' in \emph{IEEE IEDM}, 2022, pp. 23--2.

\bibitem{intelcfet}
M.~Radosavljevi{\'c} \emph{et~al.}, ``{Demonstration of a Stacked CMOS Inverter at 60nm Gate Pitch with Power Via and Direct Backside Device Contacts},'' in \emph{IEEE IEDM}, 2023, pp. 1--4.

\bibitem{tsmccfet}
S.~Liao \emph{et~al.}, ``{Complementary Field-Effect Transistor (CFET) Demonstration at 48nm Gate Pitch for Future Logic Technology Scaling},'' in \emph{IEEE IEDM}, 2023, pp. 1--4.

\bibitem{samcfet}
J.~Park \emph{et~al.}, ``{First demonstration of 3-dimensional stacked FET with top/bottom source-drain isolation and stacked n/p metal gate},'' in \emph{IEEE IEDM}, 2023, pp. 1--4.

\bibitem{ffet}
H.~Lu \emph{et~al.}, ``{First Experimental Demonstration of Self-aligned Flip FET (FFET): a Breakthrough Stacked Transistor Technology with 2.5T Design, Dual-side Active and Interconnects},'' in \emph{IEEE Symp. on VLSI Tech.}, 2024, pp. 1--2.

\bibitem{dh_cfet}
L.-C. Huang \emph{et~al.}, ``{Optimization of CFET Standard Cell Using Double-Cell-Height Structure},'' in \emph{International VLSI Symp. on Technology, Systems and Applications}, 2024, pp. 1--3.

\bibitem{cfet}
P.~Schuddinck \emph{et~al.}, ``{PPAC of sheet-based CFET configurations for 4 track design with 16nm metal pitch},'' in \emph{IEEE Symp. on VLSI Tech.}, 2022, pp. 365--366.

\bibitem{cnt7x}
C.~Gilardi \emph{et~al.}, ``{Extended scale length theory targeting low-dimensional FETs for carbon nanotube FET digital logic design-technology co-optimization},'' in \emph{IEEE IEDM}, 2021, pp. 27--3.

\bibitem{3tcfet}
S.~M.~Y. Sherazi \emph{et~al.}, ``{CFET standard-cell design down to 3Track height for node 3nm and below},'' in \emph{Design-Process-Technology Co-optimization for Manufacturability XIII}.\hskip 1em plus 0.5em minus 0.4em\relax SPIE, 2019, pp. 16--27.

\bibitem{skywater}
T.~Srimani \emph{et~al.}, ``{Foundry monolithic 3D BEOL transistor+ memory stack: Iso-performance and Iso-footprint BEOL carbon nanotube FET+ RRAM vs. FEOL silicon FET+ RRAM},'' in \emph{IEEE Symp. on VLSI Tech.}, 2023, pp. 1--2.

\bibitem{cnt_riscv}
G.~Hills \emph{et~al.}, ``Modern microprocessor built from complementary carbon nanotube transistors,'' \emph{Nature}, vol. 572, no. 7771, pp. 595--602, 2019.

\bibitem{vscnfet}
C.-S. Lee \emph{et~al.}, ``{A compact virtual-source model for carbon nanotube FETs in the sub-10-nm regime—Part II: Extrinsic elements, performance assessment, and design optimization},'' \emph{IEEE Trans. on Electron Devices}, vol.~62, no.~9, pp. 3070--3078, 2015.

\bibitem{ro}
A.~Farokhnejad \emph{et~al.}, ``{Evaluation of BEOL scaling boosters for sub-2nm using enhanced-RO analysis},'' in \emph{IEEE International Interconnect Technology Conference}, 2022, pp. 136--138.

\bibitem{gts_cd}
\emph{GTS Cell Designer 2023.9}, {Global TCAD Solutions}, 2024.

\bibitem{beol}
V.~Huang \emph{et~al.}, ``{A Comprehensive Modeling Platform for Interconnect Technologies},'' \emph{IEEE Trans. on Electron Devices}, vol.~70, no.~5, pp. 2594--2599, 2023.

\bibitem{irds}
\BIBentryALTinterwordspacing
{IEEE International Roadmap for Devices and Systems}, ``International roadmap for devices and systems 2022 edition,'' Tech. Rep., 2022. [Online]. Available: \url{https://irds.ieee.org/editions/2022}
\BIBentrySTDinterwordspacing

\bibitem{probe3}
S.~Choi \emph{et~al.}, ``{PROBE3. 0: A Systematic Framework for Design-Technology Pathfinding with Improved Design Enablement},'' \emph{IEEE Trans. on CAD of Integrated Circuits and Systems}, 2023.

\bibitem{imecpdn}
G.~Sisto \emph{et~al.}, ``{Block-level evaluation and optimization of backside PDN for high-performance computing at the A14 node},'' in \emph{IEEE Symp. on VLSI Tech.}, 2023, pp. 1--2.

\bibitem{mflowgen}
A.~Carsello \emph{et~al.}, ``{mflowgen: A modular flow generator and ecosystem for community-driven physical design},'' in \emph{Proceedings of the 59th ACM/IEEE Design Automation Conference}, 2022, pp. 1339--1342.

\bibitem{rocket}
K.~Asanovic \emph{et~al.}, ``The rocket chip generator,'' \emph{University of California, Berkeley, Tech. Rep. UCB/EECS-2016-17}, vol.~4, pp. 6--2, 2016.

\bibitem{opencore}
OpenCores, \url{https://www.opencores.org}, accessed: 2024-07-08.

\end{thebibliography}

\end{document}